# Comment on "Repulsive Casimir Force in Chiral Metamaterials"

Mário G. Silveirinha[*] and Stanislav I. Maslovski
*University of Coimbra, Department of Electrical Engineering-Instituto de Telecomunicações, 3030 Coimbra, Portugal*
(Dated: July 9, 2010)

It is shown that the proposal of Ref. [1] of Casimir repulsion and nanolevitation based on chiral metamaterials is incompatible with the passivity and causality of the materials.

As in Ref. [1], we consider an isotropic chiral metamaterial characterized by the constitutive relations: $\mathbf{D} = \varepsilon_0 \varepsilon \mathbf{E} + \frac{1}{c} i \kappa \mathbf{H}$ and $\mathbf{B} = -i\frac{1}{c}\kappa \mathbf{E} + \mu_0 \mu \mathbf{H}$. By thermodynamical considerations, the heat liberated per unit volume must be positive [2]. For harmonic electromagnetic fields with time variation $\exp(-i\omega t)$, the average heating rate in one cycle is given by ($\omega$ is real valued and positive): $q = \frac{1}{2}\mathrm{Re}\left\{-i\omega\left(\mathbf{E}^*.\mathbf{D} + \mathbf{H}^*.\mathbf{B}\right)\right\}$. Using the bianisotropic constitutive relations the heating rate may be written as a quadratic form:

$$q = \frac{\omega}{2}\begin{pmatrix}\mathbf{E}^* & \mathbf{H}^*\end{pmatrix} \cdot \begin{pmatrix} \mathrm{Im}\{\varepsilon_0 \varepsilon\} & i\mathrm{Im}\{\frac{\kappa}{c}\} \\ -i\mathrm{Im}\{\frac{\kappa}{c}\} & \mathrm{Im}\{\mu_0\mu\} \end{pmatrix} \cdot \begin{pmatrix}\mathbf{E}\\\mathbf{H}\end{pmatrix} \quad (1)$$

We used the fact that $\mathrm{Re}\{\mathbf{F}^*.\mathbf{M}.\mathbf{F}\} = \mathbf{F}^*.\mathbf{M}'.\mathbf{F}$ for a generic matrix $\mathbf{M}$ and a generic vector $\mathbf{F}$, where $\mathbf{M}' = \left(\mathbf{M} + \mathbf{M}^\dagger\right)/2$ and the $\dagger$ represents the adjoint matrix. In order that for arbitrary electric and magnetic fields $q > 0$ at a point, it is necessary that the matrix in Eq. (1) is positive definite. This ensured by the following set of necessary and sufficient conditions [3, 4]:

$$\mathrm{Im}\{\varepsilon\} > 0, \quad \mathrm{Im}\{\mu\} > 0, \quad \Delta > 0 \quad (2)$$

where $\Delta = \frac{1}{c^2}\left[\mathrm{Im}\{\varepsilon\}\mathrm{Im}\{\mu\} - (\mathrm{Im}\{\kappa\})^2\right]$ is the determinant of the matrix. The two first inequations in Eq. (2) are the familiar conditions that ensure the passivity of dielectric and magnetic materials. The third condition, $\Delta > 0$, is only relevant in case of a chiral response, since otherwise it is redundant. Clearly, it imposes a limit on the strength of the imaginary part of $\kappa$. In Fig. 1 we plot $\Delta$ as a function of frequency for the parameters used in Ref. [1]. As seen, in the case $\omega_\kappa = 0.7\omega_R$, for which the authors of Ref. [1] predicted a repulsive Casimir force, $\Delta$ can be negative, and thus such large values of the chirality parameter are incompatible with the passivity of the material [3]. Indeed, it may be checked that in order that $\Delta$ remains positive it is necessary that $\omega_\kappa < 0.032\omega_R$.

One could still try to make sense of the results of Ref. [1], and argue that for $\mathrm{Im}\{\kappa\} \approx 0$ passivity might not be violated. However, even for a lossless material the results of Ref. [1] remain unphysical. Indeed, a bianisotropic reciprocal material (such that the underlying microstructure corresponds to metal-dielectric inclusions) can be as well described by a spatially dispersive dielectric function $\bar{\bar{\varepsilon}}_{eff}(\omega, \mathbf{k})$ ($\mathbf{k}$ is the wave vector) linked to the parameters of the bianisotropic model $\varepsilon$, $\mu$ and $\kappa$ by formula (6) of Ref. [5] (see also Ref. [6]). For $\mathbf{k} = \mathbf{0}$, the nonlocal dielectric function reduces to a scalar and is given by [8]:

$$\frac{\varepsilon_{eff}}{\varepsilon_0}(\omega, \mathbf{k} = \mathbf{0}) = \varepsilon(\omega) - \frac{[\kappa(\omega)]^2}{\mu(\omega)} \quad (3)$$

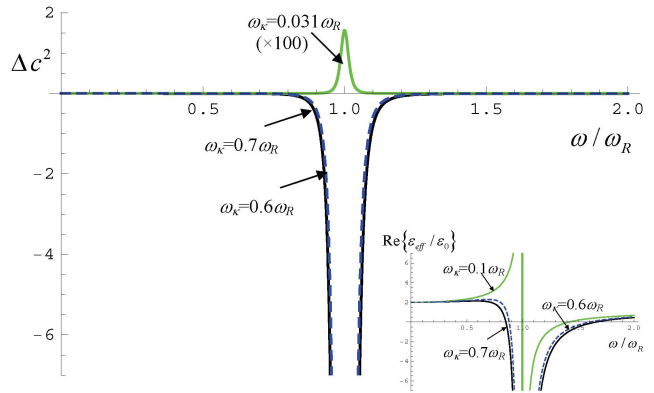

FIG. 1: (Color online) $\Delta$ as a function of $\omega/\omega_R$, for the parameters of Fig. 1 of Ref. [1]. The curves associated with $\omega_\kappa = 0.6\omega_R$ and $\omega_\kappa = 0.7\omega_R$ correspond to the parameters of the square and circle curves of Ref. [1], respectively. The inset shows $\mathrm{Re}\{\varepsilon_{eff}\}$ as a function of frequency when all parameters related to loss are set equal to zero.

The key point is that because of causality the nonlocal dielectric function satisfies for each fixed wave vector the usual Kramers-Kronig formulas (see [7] p. 14). In particular, in case of very low loss it is necessary that $\mathrm{Re}\{\varepsilon_{eff}\}$ be a strictly increasing function of frequency, i.e. $\mathrm{Re}\{\varepsilon - \kappa^2/\mu\}$ must increase with frequency (see also [3]). However, as shown in the inset of Fig. 1, when the parameters related to loss considered in [1] are set equal to zero $\mathrm{Re}\{\varepsilon_{eff}\}$ has non-monotonous behavior in case of strong chirality. Thus, the possibility of Casimir repulsion based on isotropic chiral metamaterials seems dubious.

---

# Supplementary material of the Comment "Repulsive Casimir Force in Chiral Metamaterials"

Mário G. Silveirinha, Stanislav I. Maslovski

## I. Independence of the electric and magnetic fields in a small volume

The derivation of the conditions, $\Delta > 0$, $\text{Im}\{\varepsilon\} > 0$ and $\text{Im}\{\mu\} > 0$ that ensure that $q > 0$ is based on the assumption that in a given point of space it is possible to fix the fields $\mathbf{E}$ and $\mathbf{H}$ independently. Since these fields are linked by Maxwell's equations one could argue that they cannot be independently fixed.

In this section, we prove that for an isotropic chiral material it is always possible, at least in some region of space, to make the amplitudes of the electric and magnetic fields virtually independent in a small volume with dimensions that can be some significant fraction of the wavelength. In practice, this can be achieved by using a suitable external source that encloses the region of interest. Our conclusions are consistent with those of A. Efros, who analyzed the case where magneto-electric coupling is absent [1].

We demonstrate below that it is possible to construct a solution of Maxwell's equations in the chiral material such that $\mathbf{E}$ and $\mathbf{H}$ at a given point (let's say, the origin for definiteness) are two arbitrarily specified complex vectors. Thus, to study under which conditions $q > 0$ in a vicinity of such point, we can for all purposes regard $\mathbf{E}$ and $\mathbf{H}$ as independent, and thus the analysis of our comment is physically sound.

To begin with, let us consider a plane wave of the type

$$\mathbf{E}_{\pm} = E_0 \left( \hat{\mathbf{u}}_{\phi'} \pm i\hat{\mathbf{u}}_z \right) e^{ik_{\pm}\hat{\mathbf{u}}_{\rho'} \cdot \mathbf{r}} e^{-i\omega t} \tag{A1}$$

propagating in a chiral isotropic metamaterial, where $\hat{\mathbf{u}}_{\rho'} = (\cos\phi', \sin\phi', 0)$ is the direction of propagation (determined by the angle $\phi'$), $\hat{\mathbf{u}}_{\phi'} = (-\sin\phi', \cos\phi', 0)$ and $\hat{\mathbf{u}}_z = (0,0,1)$. The propagation constant is:

$$k_{\pm} = \frac{\omega}{c}\left(\sqrt{\mu\varepsilon} \pm \chi\right) \tag{A2}$$

and the magnetic field associated with $\mathbf{E}_{\pm}$ is such that:

$$\mathbf{H}_{\pm} = -\frac{1}{\eta_0 z}(\pm i)\mathbf{E}_{\pm} \tag{A3}$$

being $z = \mu/n = \mu/\sqrt{\mu\varepsilon}$ the normalized impedance, and $\eta_0 = \sqrt{\mu_0/\varepsilon_0}$ the intrinsic impedance of vacuum. Since Maxwell's equations are linear we can construct other solutions by superposing plane waves associated with different values of $\phi'$. For example, we can consider the solutions:

$$\mathbf{E}_{\pm,c}(\mathbf{r}) = \frac{1}{2\pi}\int_0^{2\pi} \mathbf{E}_{\pm}(\mathbf{r};\phi')d\phi'. \tag{A4}$$

Writing the observation point $\mathbf{r}$ in cylindrical coordinates $(\rho,\phi)$, it follows that:

$$\mathbf{E}_{\pm,c}(\mathbf{r}) = E_0 e^{-i\omega t}\frac{1}{2\pi}\int_0^{2\pi}(-\sin\phi', \cos\phi', \pm i)e^{ik_{\pm}\rho\cos(\phi'-\phi)}d\phi'. \tag{A5}$$

After straightforward calculations, this yields:

$$\mathbf{E}_{\pm,c}(\mathbf{r}) = iE_0 e^{-i\omega t}\left(\hat{\mathbf{u}}_{\phi}J_1(k_{\pm}\rho) \pm \hat{\mathbf{u}}_z J_0(k_{\pm}\rho)\right) \tag{A6}$$

$$\mathbf{H}_{\pm,c}(\mathbf{r}) = H_0 e^{-i\omega t}\left(\pm\hat{\mathbf{u}}_{\phi}J_1(k_{\pm}\rho) + \hat{\mathbf{u}}_z J_0(k_{\pm}\rho)\right)$$

where $J_n$ is the Bessel function of first kind and order $n$, and $H_0 = E_0/(\eta_0 z)$.

In particular, let us define the following functions:

$$\mathbf{E}_{1,z}(\mathbf{r}) = \frac{\mathbf{E}_{+,c}(\mathbf{r}) + \mathbf{E}_{-,c}(\mathbf{r})}{2}; \qquad \mathbf{E}_{2,z}(\mathbf{r}) = \frac{\mathbf{E}_{+,c}(\mathbf{r}) - \mathbf{E}_{-,c}(\mathbf{r})}{2} \tag{A7}$$

It is evident that both $\mathbf{E}_{1,z}$ and $\mathbf{E}_{2,z}$ are solutions of Maxwell's equations in the chiral material, and it can be checked that:

$$\mathbf{E}_{1,z}(\mathbf{r}=0) = 0 \quad ; \quad \mathbf{H}_{1,z}(\mathbf{r}=0) = H_1\hat{\mathbf{u}}_z \neq 0 \tag{A8}$$

$$\mathbf{E}_{2,z}(\mathbf{r}=0) = E_2 \hat{\mathbf{u}}_z \neq 0 \quad ; \quad \mathbf{H}_{2,z}(\mathbf{r}=0) = 0 \tag{A9}$$

By symmetry it is possible to construct functions $\mathbf{E}_{1,x}$, $\mathbf{E}_{1,y}$, $\mathbf{E}_{2,x}$, $\mathbf{E}_{2,y}$ with properties analogous to those of $\mathbf{E}_{1,z}$ and $\mathbf{E}_{2,z}$, except that in Eqs. (A8)-(A9) the unit vector $\hat{\mathbf{u}}_z$ is replaced by either $\hat{\mathbf{u}}_x$ or $\hat{\mathbf{u}}_y$.

Thus, by considering a suitable linear combination of $\mathbf{E}_{1,x}$, $\mathbf{E}_{1,y}$, $\mathbf{E}_{2,x}$, $\mathbf{E}_{2,y}$, $\mathbf{E}_{1,z}$ and $\mathbf{E}_{2,z}$, we can always synthesize a field distribution that satisfies the homogeneous Maxwell's equations in the chiral material with completely arbitrary values of $\mathbf{E}$ and $\mathbf{H}$ (not restricted in any manner) in the vicinity of the origin.

It is interesting to note that in presence of loss the considered electromagnetic field distribution can only exist in a bounded region of space (because $|J_n(k_\pm \rho)| \to \infty$ when $\rho \to \infty$, if $k_\pm$ is complex valued). This means that the electromagnetic field distribution must be created by some external source enclosing the volume of interest.

Finally, we would like to give an additional and very simple proof that the parameters considered in Ref. [2] are indeed unphysical (or at least incompatible with the requirement of passivity). It is sufficient to consider the simple problem, where a plane wave illuminates along the normal direction a chiral slab standing in vacuum. Let us assume that the incoming plane wave is right circularly polarized (RCP polarization). The transmission coefficient can be calculated using formula (13) of Ref. [3], which is repeated below for convenience:

$$T_+ = \frac{4z e^{ik_+ d}}{(1+z)^2 - (1-z)^2 e^{2ink_0 d}} \tag{A10}$$

In the above $k_+ = n + \chi$, $n = \sqrt{\mu\varepsilon}$, and $z = \mu/n = \sqrt{\mu/\varepsilon}$ is a normalized impedance. Using the above formula we have calculated the amplitude of the transmission coefficient as a function of frequency for different values of the chirality parameter $\omega_\kappa$, and for the same parameters as in Fig. 1 of our Comment. In all the simulations we fixed arbitrarily the thickness of the slab $d$ such that $\frac{\omega_R}{c}d = 1.0$ ($\omega_R$ is a normalization parameter used in the material parameters of Ref. [2]).

The obtained results are plotted in Fig. 1 shown below. It is seen that when the chirality parameter $\omega_\kappa$ is large ($\omega_\kappa = 0.2\omega_R$ and $\omega_\kappa = 0.7\omega_R$) the amplitude of the transmission coefficient can exceed unity. This implies unquestionably that the material is active. The transmission coefficient is less than unity only when the chirality parameter is sufficiently small (e.g. $\omega_\kappa = 0.031\omega_R$), consistent with the theoretical analysis of our Comment.

We would like to point out that when using Eq. (A10) one must choose the square root branches used in the definitions of $n = \sqrt{\mu\varepsilon}$ and $z = \sqrt{\mu/\varepsilon}$ consistently. Specifically, the correct definition of $z$ is such that $z = \mu/n$. With such choice formula (A10) is invariant independently of the choice of the square root branch in $n = \sqrt{\mu\varepsilon}$.

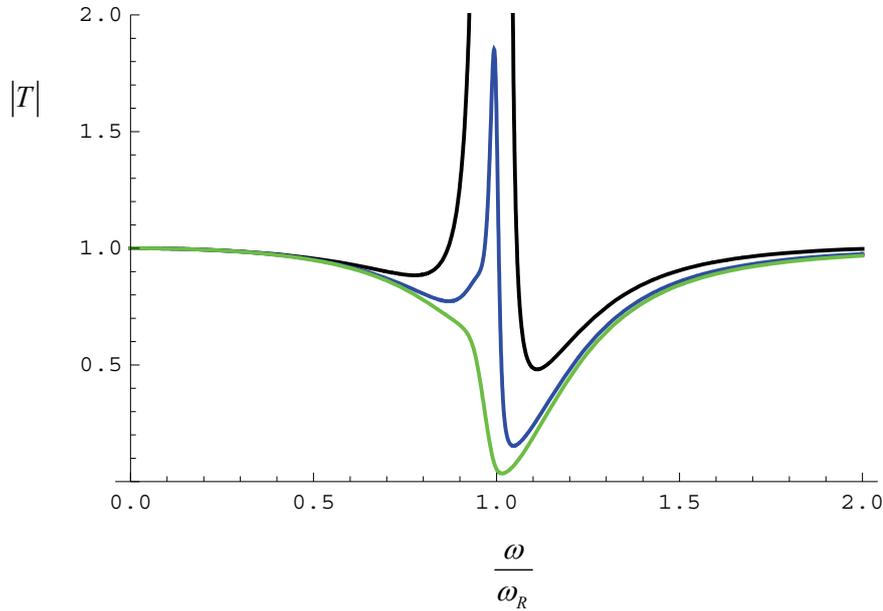

**Fig. A1** Amplitude of the transmission coefficient as a function of the normalized frequency and different values of the chirality parameter $\omega_\kappa$. Black line: $\omega_\kappa = 0.7\omega_R$; Blue line: $\omega_\kappa = 0.2\omega_R$; Green line: $\omega_\kappa = 0.031\omega_R$.

# II. On the restrictions on $\varepsilon - \chi^2/\mu$ for low loss materials

It is interesting to note that the parameter $\varepsilon - \chi^2/\mu$ considered in our Comment has a quite simple physical meaning if one considers an alternative set of constitutive relations also used very often in the analysis of chiral media.

In fact, there are two popular (and equivalent) sets of constitutive relations (among many other possibilities) that are used to study chiral materials. The first set is the one used in the letter of Zhao, *et al* (Ref. [2]), which reads:

$$\mathbf{D} = \varepsilon_0 \varepsilon \mathbf{E} + \frac{i\kappa}{c}\mathbf{H} \qquad \mathbf{B} = -\frac{i\kappa}{c}\mathbf{E} + \mu_0 \mu \mathbf{H} \qquad (B1)$$

This set of constitutive relations puts emphasis on the $\mathbf{E}$ and $\mathbf{H}$ fields. However, since $\mathbf{H}$ is a derived field, $\mathbf{H} \equiv \mathbf{B}/\mu_0 - \mathbf{M}$ some authors prefer to put emphasis on the $\mathbf{E}$ and $\mathbf{B}$ fields, i.e. on the macroscopic version of the microscopic fields. This yields the alternative set of constitutive relations:

$$\mathbf{D} = \varepsilon_0 \varepsilon' \mathbf{E} + \frac{i\kappa'}{c\mu_0}\mathbf{B} \qquad \mathbf{H} = \frac{i\kappa'}{c\mu_0}\mathbf{E} + \mu_0^{-1}\mu'^{-1}\mathbf{B} \qquad (B2)$$

In order that the two sets of constitutive relations are equivalent it is necessary that the primed and unprimed parameters are related as follows:

$$\varepsilon' = \varepsilon - \frac{\kappa^2}{\mu}, \qquad \kappa' = \frac{\kappa}{\mu}, \qquad \mu' = \mu \qquad (B3)$$

Thus, within the framework of the constitutive relations (B2), the parameter $\varepsilon - \chi^2/\mu$ considered in our comment is nothing more than the "permittivity" $\varepsilon'$ of the model (B2). In particular, our conclusion that in the case of low loss $\varepsilon - \chi^2/\mu$ must be an increasing function frequency, is equivalent to say that in case of low loss $\varepsilon'$ must be an increasing function of frequency (as expected from the usual Kramers-Kronig's relations [4]).